\documentclass[12pt]{article}

\usepackage{abstract} 
\usepackage{amsfonts,amsmath,amssymb,amsthm}
\usepackage{array}
\usepackage[superscript,biblabel]{cite}
\usepackage[mathscr]{eucal}
\usepackage{float}
\usepackage{latexsym}
\usepackage{setspace}
\usepackage{sectsty}
	\sectionfont{\large}
	\subsectionfont{\normalsize}
\usepackage{slashbox}
\usepackage{soul}
\usepackage{tabularx}
\usepackage{multirow}
\usepackage[lofdepth,lotdepth]{subfig}
\usepackage[margin=2cm,font=footnotesize]{caption}
\usepackage{titling}
\usepackage{bm}
\newcommand{\subtitle}[1]{%
  \posttitle{%
    \par\end{center}
    \begin{center}\large#1\end{center}
   \vskip0.1cm}%
    }
\renewcommand{\arraystretch}{1.2}
\usepackage[final]{graphicx}   
 
\numberwithin{equation}{section}
\usepackage{stackengine}[2013-09-11]%
\def\refA{$m$}%
\def\refB{$m^{\prime\prime}$}%
\newcommand\bsbox[2]{%
  \raisebox{-1pt}{%
  \stackinset{c}{}{c}{}{\rotatebox{45}{\rule[-.5em]{.2pt}{3em}}}{%
  \def\stacktype{L}%
  \setstackgap{L}{.5\baselineskip}%
  \kern-1pt\makebox[\widthof{\refA}][r]{#1}\stackon{}{\smash{\makebox[\widthof{\refB}]{#2}}}\kern1pt%
  }%
  \kern-5.5pt}%
}
\usepackage[margin=1cm,font=footnotesize]{caption}
\usepackage[margin=3cm,columnsep=20pt]{geometry}

\begin{document}

\setlength{\parindent}{1cm}
\setlength{\belowdisplayskip}{5pt} \setlength{\belowdisplayshortskip}{5pt}
\setlength{\abovedisplayskip}{5pt} \setlength{\abovedisplayshortskip}{5pt}

\theoremstyle{definition}
\newtheorem{defn}{Definition}
\newtheorem{example}{Example}
\newtheorem{theorem}{Theorem}
\newtheorem{Prob}{Problem}

\title{\Large{\vspace{-10ex}The Preon Sector of the SLq(2) Model\\and the Binding Problem\vspace{-2ex}}}
\author{\normalsize{$Robert J. Finkelstein$}  \\ \normalsize{Department of Physics \& Astronomy\vspace{-2ex}}  \\ \normalsize{University of California, Los Angeles, 90095-1547}}
\date{}


\maketitle
\thispagestyle{empty}

\begin{abstract}{\normalsize{There are suggestive experimental indications that the leptons, neutrinos, and quarks might be composite and that their structure is described by the quantum group SLq(2). Since the hypothetical preons must be very heavy relative to the masses of the leptons, neutrinos, and quarks, there must be a very strong binding field to permit these composite particles to form. Unfortunately there are no experiments direct enough to establish the order of magnitude needed to make the SLq(2) Lagrangian dynamics quantitative. It is possible, however, to parametrize the preon masses and interactions that would be necessary to stabilize the three particle composite representing the leptons, neutrinos, and quarks. In this note we examine possible parametrizations of the masses and the interactions of these hypothetical structures. We also note an alternative view of SLq(2) preons.}}
\end{abstract}

\newpage
\setcounter{page}{1}

\section{Introduction}
There is suggestive experimental support for the view that the leptons, neutrinos, and quarks have a preonic substructure\cite{art1,art2} and that this substructure is described by the knot algebra SLq(2)\cite{art3}. 
In the SLq(2) model the preons are described by the fundamental representation, and the vector bosons, by which the preons interact, are described by the adjoint representation of SLq(2). Denoting the elements of the irreducible representations by D$^j_{m m^\prime}$, 
the fundamental representation is denoted by $j=\frac 1 2$, and the adjoint representation by $j=1$. In the same model, the elementary fermions of the standard 
model, i.e., the charged leptons, neutrinos, and quarks, lie in the $j=\frac 3 2$ representation and the electroweak vectors in the $j=3$ representation. In a more detailed description, the elementary fermions of the standard model are composed of three $j=\frac 1 2$ preons bound by the $j=1$ adjoint vectors. Since the masses of the composite particles are relatively very light compared to the masses of the heavy preons, the model requires a very strong binding force. The natural agent for this binding is the preonic vector field, which is, in this model, solely responsible for the preon-preon interaction. There are other candidates for the role of the binding force, but these are external to the knot model\cite{art4}. Since the composite preonic particles must be so small and the preons so heavy, however, it is possible that gravitational attraction would also play a significant role.


\section{Representation of SLq(2)}
The two-dimensional representation of SLq(2) may be defined as follows:
\begin{equation}
\mbox T=\mbox D^{\frac 1 2}_{m m^\prime}=\left( \def\arraystretch{1}\begin{array}{c c}
a & b \\
c & d \end{array} \right),
\end{equation}
where ($a$,$b$,$c$,$d$) satisfy the knot algebra:
\begin{equation*}
\begin{aligned}[c]
ab&=qba\\
ac&=qca
\end{aligned}
\qquad \qquad
\begin{aligned}[c]
bd&=qdb\\
cd&=qdc
\end{aligned}
\qquad \qquad
\begin{aligned}[c]
ad-qbc&=1\\
da-q_1cb&=1
\end{aligned}
\qquad \qquad
\begin{aligned}[c]
bc&=cb\\
q_1&\equiv q^{-1}.
\end{aligned} \tag{A}
\end{equation*}
The following two-dimensional matrix
\begin{equation}
\varepsilon=\left(\def\arraystretch{1} \begin{array}{c c}
0 & \alpha_2 \\
-\alpha_1 & 0 \end{array} \right)
\end{equation}
is invariant under
\begin{equation}
\mbox T\varepsilon \mbox T^t=\mbox T^t\varepsilon \mbox T=\varepsilon,
\end{equation}
where T$^t$ is T transposed, and
\begin{equation*}
\frac{\alpha_1}{\alpha_2}=q.
\end{equation*}

To obtain the higher representations of SLq(2), we transform the $(2j+1)$ monomials\cite{art5},
\begin{equation}
\begin{aligned}[c]
\Psi^j_m=\mbox N^j_m x^{n_+}_1 x^{n_-}_2,
\end{aligned}
\qquad 
\begin{aligned}[c]
-j \le m \le j
\end{aligned}
\end{equation}
by
\begin{equation}
\left( \def\arraystretch{1}\begin{array}{c}
x_1 \\
x_2 \end{array} \right)^\prime=\mbox T\left( \def\arraystretch{1}\begin{array}{c}
x_1 \\
x_2 \end{array} \right),
\end{equation}
or
\begin{align}
x_1^\prime&=a x_1+b x_2 \\
x_2^\prime&=cx_1+dx_2,
\end{align}
where ($a$,$b$,$c$,$d$) satisfy the knot algebra, but $x_1$ and $x_2$ commute:
\begin{equation}
[x_1,x_2]=0.
\end{equation}
Here
\begin{equation}
n_\pm=j\pm m,
\end{equation}
and
\begin{equation}
\mbox N^j_m=\left[\langle n_+\rangle_{q_1}!\langle n_-\rangle_{q_1}!\right]^{-\frac 1 2},
\end{equation}
where
\begin{equation}
\langle n\rangle_q=\frac{q^n-1}{q-1}.
\end{equation}
One finds\cite{art5}
\begin{equation}
\left(\Psi^j_m\right)^\prime=\sum \mbox D^j_{mm^\prime}\Psi^j_{m^\prime},
\end{equation}
where
\begin{equation}
\begin{aligned}[c]
\mbox D^j_{mm^\prime}\left(q | a,b,c,d\right)=&~~~\sum ~~~\mbox A^j_{mm^\prime}(q,n_a,n_c)\delta(n_a+n_c,n^\prime_+)a^{n_a}b^{n_b}c^{n_c}d^{n_d} \\[-1em]
&\scriptstyle{\delta(n_a+n_b,n_+)}\\[-1.2em]
&\scriptstyle{\delta(n_c+n_d,n_-)}
\end{aligned}
\end{equation}
where the sum is over the positive integers ($n_a,n_b,n_c,n_d$) subject to the $\delta$-function constraints as shown.

\noindent Here
\begin{equation}
\mbox A^j_{mm^\prime}(q,n_a,n_c)=\left[\frac{\langle n^\prime_+\rangle_1! \langle n^\prime_-\rangle_1!}{\langle n_+\rangle_1!\langle n_-\rangle_1!}\right]^{\frac 1 2} \frac{\langle n_+\rangle_1!}{\langle n_a\rangle_1!\langle n_b\rangle_1!}\mbox{  } \frac{\langle n_-\rangle_1!}{\langle n_c\rangle_1!\langle n_d\rangle_1!},
\end{equation}
where
\begin{equation}
\begin{aligned}[c]
n_\pm&=j\pm m\\
n^\prime_\pm&=j\pm m^\prime,
\end{aligned}
\end{equation}
and
\begin{equation}
\langle n\rangle_1=\frac{q^n_1-1}{q_1-1}.
\end{equation}
We take $q$ to be real.

\noindent The algebra (A) is invariant under the gauge transformations:
\begin{equation}
\begin{aligned}[c]
\mbox U_a(1): \mbox{  }\\
\mbox{ }
\end{aligned} 
\begin{aligned}[c]
a^\prime&=e^{\imath \varphi_a}a\\
d^\prime&=e^{-\imath \varphi_a}d
\end{aligned}
\qquad \qquad
\begin{aligned}[c]
\mbox U_b(1):  \mbox{  }\\
\mbox{ }
\end{aligned} 
\begin{aligned}[c]
b^\prime&=e^{\imath \varphi_b}b\\
c^\prime&=e^{-\imath \varphi_b}c
\end{aligned}
\end{equation}
Then U$_a(1)\times \mbox U_b(1)$ induces on D$^j_{mp}(a,b,c,d)$ the gauge transformation
\begin{equation}
\mbox D^j_{mp}(a^\prime,b^\prime,c^\prime,d^\prime)=e^{\imath (\varphi_a+\varphi_b)m}e^{\imath(\varphi_a-\varphi_b)p}\mbox { D}^j_{mp}(a,b,c,d),
\end{equation}
or
\begin{equation}
{\mbox D^j_{mp}}^\prime=\mbox U_m \times \mbox U_p \mbox{ D}^j_{mp}
\end{equation}


\section{The SLq(2) Extension of the Standard Model and the \\Quantum Knot}
To obtain the SLq(2) extension of the standard model, replace the field operators, $\Psi$, of the standard model by
\begin{equation}
\hat\Psi^j_{mp}=\Psi\mbox D^j_{mp}(\Psi),
\end{equation}
where the $\Psi$ may be an elementary fermion, weak vector, or Higgs field operator. The D$^j_{mp}$ are elements of the irreducible representations of the knot algebra SLq(2). The normal modes of $\hat\Psi^j_{mp}$ define the field quanta of the extended model, and these field quanta will be called ``quantum knots."

We postulate a correspondence between quantum knots and oriented classical knots according to
\begin{equation}
(j,m,p)=\frac 1 2 (N,w,r+o),
\end{equation}
where $(N,w,r)$ are (the number of crossings, the writhe, and the rotation, respectively) of the 2d-projection of an oriented classical knot. Since the $(N,w,r)$ are integers, the factor $\frac 1 2$ is needed to allow half-integer representations of SLq(2). Since $2m$ and $2p$ are of the same parity, while $w$ and $r$ are topologically constrained to be of opposite parity, $o$ is an odd integer that we set $=1$ for a quantum trefoil.

Equation (3.2) restricts the states of the quantum knot to only those states of the full $2j+1$ dimensional representations that correspond to the 2d-spectrum $(N,w,r)$ of a corresponding oriented classical knot. The algebra (A) and the representation (2.13) make no references to orientation. By imposing (3.2) we are relating (2.13) to oriented knots. This turns out to be important for the physical interpretation.

The defining SLq(2) algebra is invariant under gauge transformations 
U$_a(1)\times \mbox U_b(1)$ that induce the gauge transformations (2.19) on the D$^j_{mp}$\cite{art5,art6} and hence on the field operators $\hat\Psi^j_{mp}$:
\begin{equation}
\mbox{$\hat\Psi^j_{mp}$}^\prime=\mbox U_m(1) \times \mbox U_p(1) \hat\Psi^j_{mp}.
\end{equation}
For physical consistency the field action is required to be invariant under Equation (3.3) since the U$_a(1)\times \mbox U_b(1)$ transformations do not change the defining algebra.

Then there will be the following Noether charges that may be described by (3.2) as writhe and rotation charges
\begin{align}
Q_w&\equiv -k_wm=-k_w\frac w 2\\
Q_r&\equiv -k_rp=-k_r\frac{r+o} 2
\end{align}

We assume that $k=k_w=k_r$ is a universal constant with the dimensions of an electric charge and with the same value for all trefoils.

The knot picture is more plausible if the simplest particles are the simplest knots. We therefore consider the possibility that the most elementary fermions with isotopic spin $t=\frac 1 2$ are the most elementary quantum knots, the quantum trefoils with $N=3$. This possibility is supported by the following empirical observation
\begin{equation}
(t,-t_3,-t_0)=\frac 1 6 (N,w,r+1),
\end{equation}
which is satisfied by the four classes of elementary fermions described by $(\frac 1 2,t_3,t_0)$ and the four quantum trefoils described by $(3,w,r)$ and shown by the row-to-row correspondence in Table 3.1.

\renewcommand\arraystretch{1.7}
\begin{table}[ht]
\centering
{\textbf{\ul{Table 3.1}}}\\[1ex]
$\begin{array}{c c rr |c c rr}\hline\hline
(f_1,f_2,f_3) & t & t_3 & t_0 & \mbox D^{\frac N 2}_{\frac w 2 \frac{r+1}2} & N & w & r \\ [1ex]\hline
(e,\mu,\tau)_L & \frac 1 2 & -\frac 1 2 & -\frac 1 2 & \mbox D^{\frac 3 2}_{\frac 3 2 \frac 3 2} & 3 & 3 & 2 \\
(\nu_e,\nu_\mu,\nu_\tau)_L & \frac 1 2 & \frac 1 2 & -\frac 1 2 & \mbox D^{\frac 3 2}_{-\frac 3 2 \frac 3 2} & 3 & -3 & 2\\
(d,s,b)_L & \frac 1 2 & -\frac 1 2 & \frac 1 6 &  \mbox D^{\frac 3 2}_{\frac 3 2 -\frac 3 2} & 3 & 3 & -2 \\
(u,c,t)_L & \frac 1 2 & \frac 1 2 & \frac 1 6 & \mbox D^{\frac 3 2}_{-\frac 3 2 -\frac 3 2} & 3 & -3 & -2 \\[1ex]\hline\hline
\end{array}$
\end{table}

Only for the particular row-to-row correspondence shown in Table 3.1 does (3.6) hold, i.e., each class of fermions $(t_3,t_0)$ is \ul{uniquely correlated with a specific $(w,r)$ trefoil, and therefore with a specific D$^{\frac 3 2}_{mm^\prime}$}.

By (3.2) and (3.6) one also has
\begin{equation}
(j,m,m^\prime)=3(t,-t_3,-t_0)
\end{equation}
for the fermion quantum trefoils.

In the knot model quantum knots are jointly defined by the topological condition (3.2) and the empirical constraint (3.7).

\renewcommand\arraystretch{1.5}
\begin{table}[H]
\hspace*{-0.25in}
\centering
{\textbf{\ul{Table 3.2}}}
\resizebox{\columnwidth}{!}{%
\small{
\begin{tabular}{c r r r c | c l c c c}
\multicolumn{5}{c }{Standard Model} & \multicolumn{5}{c}{Quantum Trefoil Model}\\ \hline \hline
$(f_1,f_2,f_3)$ & $t$ & $t_3$ & $t_0$ & $Q_e$ & $(w,r)$ & D$^{\frac N 2}_{\frac w 2 \frac{r+1}2}$ & $Q_w$ & $Q_r$ & $Q_w+Q_r$ \\[1ex] \hline
$(e,\mu,\tau)_L$ & $\frac 1 2$ & $-\frac 1 2$ & $-\frac 1 2$ & $-e$ & (3,2) & D$^{\frac 3 2}_{\frac 3 2 \frac 3 2}$ & $-k\left(\frac 3 2\right)$ & $-k\left(\frac 3 2\right)$ & $-3k$ \\[2ex]
$(\nu_e,\nu_\mu,\nu_\tau)_L$ & $\frac 1 2$ & $\frac 1 2$ & $-\frac 1 2$ & 0 & (-3,2) & D$^{\frac 3 2}_{-\frac 3 2 \frac 3 2}$ & $-k\left(-\frac 3 2\right)$ & $-k\left(\frac 3 2\right)$ & 0\\[2ex]
$(d,s,b)_L$ & $\frac 1 2$ & $-\frac 1 2$ & $\frac 1 6$ & $-\frac 1 3 e$ & (3,-2) & D$^{\frac 3 2}_{\frac 3 2 -\frac 1 2}$ & $-k\left(\frac 3 2\right)$ & $-k\left(-\frac 1 2\right)$ & $-k$ \\[2ex]
$(u,c,t)_L$ & $\frac 1 2$ & $\frac 1 2$ & $\frac 1 6$ & $\frac 2 3 e$ & (-3,-2) & D$^{\frac 3 2}_{-\frac 3 2 -\frac 1 2}$ & $-k\left(-\frac 3 2\right)$ & $-k\left(-\frac 1 2\right)$ & $2k$ \\[1ex]
& & & & $Q_e=e(t_3+t_0)$ & & & $Q_w=-k\frac w 2$ & $Q_r=-k\frac{r+1}2$ & \\\hline\hline
\end{tabular}}%
}
\end{table}

In Table 3.2 we next compare the charges $Q_e$ of the observed fermions with the total charges of the quantum trefoils. To construct and interpret this table we have postulated that $k=k_w=k_r$ is a universal constant with the same value for all trefoils. We then obtain the value of $k$ by requiring that the total charge, $Q_w+Q_r$, of each quantum trefoil satisfy
\begin{equation}
Q_e=Q_w+Q_r,
\end{equation}
where $Q_e$ is the electric charge of the corresponding family of elementary fermions as shown in Table 3.2.

\noindent Then
\begin{equation}
k=\frac e 3,
\end{equation}
and $t_3$ and $t_0$ measure the writhe and rotation charges:
\begin{align}
Q_w&=et_3\left(=-\frac e 3 m=-\frac e 6 w\right)\\
Q_r&=et_0\left(=-\frac e 3 m^\prime=-\frac e 6 (r+1)\right)
\end{align}
Then by (3.8), (3.10), and (3.11),
\begin{equation}
Q_e=e(t_3+t_0),
\end{equation}
and by (3.10) and (3.11)
\begin{equation}
Q_e=-\frac e 3 (m+m^\prime),
\end{equation}
or
\begin{equation}
Q_e=-\frac e 6 (w+r+1).
\end{equation}

Then the electric charge is a measure of the total writhe + rotation, of the trefoil. The total electric charge in this way resembles the total angular momentum as a sum of two parts where the localized contribution of the writhe to the charge corresponds to the localized contribution of the spin to the angular momentum.

We consider only quantum knots that carry the charge expressed as both (3.12) and (3.13).


\section{The SLq(2) Extension of the Standard Model}
One may give physical meaning to the defining expression (2.13) for D$^j_{mm^\prime}$ by postulating that  $(a,b,c,d)$ are  creation operators for fermionic preons. Then the elements of the fundamental $(j=\frac 1 2)$ representations may be interpreted as creation operators for the four preons shown in (4.1).
\begin{equation}
\mbox D^{\frac 1 2}_{mm^\prime}=\arraycolsep=5pt\def\arraystretch{1}\begin{array}{r | r r}
\mbox{\normalsize\bsbox{$m$}{$m^{\prime\prime}$}}& \frac 1 2 & -\frac 1 2 \\ \hline
\frac 1 2 & a & b \\
-\frac 1 2 & c & d
\end{array}
\end{equation}

By (4.1) and (3.13) there is \ul{one charged preon, $a$, with charge $-\frac e 3$ and its antiparticle, $d$, and there is one neutral particle, $b$, with its antiparticle, $c$}.

By (3.2) the corresponding $(a,b,c,d)$ classical configurations cannot be described as knots since they have only a single crossing. They can, however, be interpreted as twisted loops with $w=\pm1$ and $r=0$. We shall give a physical meaning to these twisted loops by interpreting them as flux tubes, and we shall regard $a,b,c,d$ as creation operators for either preonic particles or preonic flux tubes, depending on whether they concentrate energy and momentum at a point or on a curve.

Then every D$^j_{mm^\prime}$ as given in (2.13), being a polynomial in $a,b,c,d$, can be interpreted as creating a superposition of states, each state with $n_a,n_b,n_c,n_d$ preons. The $(a,b,c,d)$ population of each of these states is constrained by the triplet $(j,m,m^\prime)$ that allows $(n_a,n_b,n_c,n_d)$ to vary but fixes $(t,t_3,t_0)$ and $(N,w,r+o)$ according to (3.7) and (3.2).

It then turns out that the creation operators for the leptons, D$^{\frac 3 2}_{\frac 3 2 \frac 3 2}$, neutrinos, D$^{\frac 3 2}_{-\frac 3 2 \frac 3 2}$, down quarks, D$^{\frac 3 2}_{\frac 3 2 -\frac 1 2}$, and up quarks, D$^{\frac 3 2}_{-\frac 3 2 -\frac 1 2}$, as required by Tables 3.1 and 3.2, are represented by (2.13) as the following monomials
\begin{equation}
\mbox D^{\frac 3 2}_{\frac 3 2 \frac 3 2}\sim a^3, 
\qquad \mbox D^{\frac 3 2}_{-\frac 3 2 \frac 3 2}\sim c^3,
\qquad \mbox D^{\frac 3 2}_{\frac 3 2 -\frac 1 2}\sim a b^2,
\qquad \mbox D^{\frac 3 2}_{-\frac 3 2 -\frac 1 2}\sim c d^2
\end{equation}
so that \ul{leptons and neutrinos are composed of three $a$-preons and three $c$-preons, respectively, while the down quarks are composed of one $a$- and two $b$-preons, and the up quarks are composed of one $c$- and two $d$-preons}. Both (4.1), with (3.13), and (4.2) are in agreement with the Harari-Shupe model.

The previous considerations are based on electroweak physics. To describe the strong 
interactions it is necessary according to the standard model 
to introduce SU(3). In the SLq(2) electroweak model, as here described, 
the need for the additional SU(3) symmetry appears already at the level of 
the charged leptons and neutrinos since they are presented in the SLq(2) model as $a^3$ and $c^3$, respectively. Then the simple way to protect the Pauli principle is to make the replacements
\begin{align*}
\mbox{leptons:  }& \mbox{ } a^3 \rightarrow \varepsilon^{ijk}a_ia_ja_k \\
\mbox{neutrinos:  }&  \mbox{ } c^3 \rightarrow \varepsilon^{ijk}c_ic_jc_k
\end{align*}
where $a_i$ and $c_i$ provide a basis for the fundamental representation of SU(3). Then the leptons and neutrinos are color singlets. If the $b$ and $d$ preons are also color singlets, then down quarks $a_ib^2$ and up quarks $c_id^2$ provide a basis for the fundamental representation of SU(3), as required by the standard model.


\section{Complementarity}
The representation of D$^j_{mm^\prime}$ as a function of $(a,b,c,d)$ and $(n_a,n_b,n_c,n_d)$ by Equation (2.13) implies the following constraints on the exponents:
\begin{align}
n_a+n_b+n_c+n_d &= 2j\\
n_a+n_b-n_c-n_d &= 2m\\
n_a-n_b+n_c-n_d &= 2m^\prime
\end{align}
The two relations giving physical meaning to D$^j_{mm^\prime}$, namely
\begin{equation}
(j,m,m^\prime)=\frac 1 2(N,w,r+o)
\end{equation}
and
\begin{equation}
(j,m,m^\prime)=3(t,-t_3,-t_0)
\end{equation}
imply two different interpretations of the relations (5.1)-(5.3). By (5.4) one has
\begin{align}
N &= n_a+n_b+n_c+n_d\\
w &= n_a+n_b-n_c-n_d
\end{align}\begin{align}
\tilde r &= n_a-n_b+n_c-n_d
\end{align}
and by (5.5) one has
\begin{align}
t &= \frac 1 6(n_a+n_b+n_c+n_d)\\
t_3 &= -\frac {1}6 (n_a+n_b-n_c-n_d)\\
t_0 &= -\frac 1 6 (n_a-n_b+n_c-n_d)
\end{align}
In (5.8), $\tilde r \equiv r+o$.

These relations hold for all representations allowed by the model. For the fundamental representation they imply the Tables 5.1 and 5.2 describing the fermionic preons.

\renewcommand\arraystretch{1.2}
\begin{figure}[h]
\captionsetup[subfigure]{labelformat=empty}\centering
 \subfloat[][]{
\begin{tabular}{l | r r r}
\multicolumn{4}{c}{\ul{\textbf{Table 5.1}}}\\ [2ex] \hline\hline
$p$ & $N_p$ & $w_p$ & $\tilde r_p$ \\ [1ex] \hline 
$a$ & $1$ & $1$ & $1$ \\[1ex]
$b$ & $1$ & $1$ & $-1$ \\[1ex]
$c$ & $1$ & $-1$ & $1$ \\[1ex]
$d$ & $1$ & $-1$ & $-1$ \\[1ex]\hline\hline
\end{tabular}}
\qquad\qquad\qquad
 \subfloat[][]{
\begin{tabular}{l |r r r r}
\multicolumn{5}{c}{\ul{\textbf{Table 5.2}}}\\ [2ex]\hline\hline
$p$ & $t_p$ & $t_{3_p}$ & $t_{0_p}$ & $Q_p$ \\[1ex]\hline
$a$ & $\frac 1 6$ & $-\frac 1 6$ & $-\frac 1 6$ & $-\frac e 3$ \\[1ex]
$b$ & $\frac 1 6$ & $-\frac 1 6$ & $\frac 1 6$ & 0\\[1ex]
$c$ & $\frac 1 6$ & $\frac 1 6$ & $-\frac 1 6$ & 0 \\[1ex]
$d$ & $\frac 1 6$ & $\frac 1 6$ & $\frac 1 6$ & $\frac e 3$ \\[1ex]\hline\hline
\end{tabular}}
\end{figure}

\noindent By Equations (5.6)-(5.8) and Table 5.1,
 \begin{align}
 N&=\sum_p n_p N_p\\
 w&=\sum_p n_p w_p, \qquad\qquad p=(a,b,c,d)\\
 \tilde r&=\sum_p n_p \tilde r_p
 \end{align}
and by Equations (5.9)-(5.11) with Table 5.2,
 \begin{align}
 t&=\sum_p n_p t_p\\
 t_3&=\sum_p n_p t_{3_p}, \qquad\qquad p=(a,b,c,d).\\
t_0 &=\sum_p n_p t_{0_p}
 \end{align}
Here we have introduced the ``quantum rotation" $\tilde r$:
\begin{equation*}
\tilde r=r+o. 
\end{equation*}
Since $r=0$ for preons,
\begin{equation}
\tilde r_p=o_p.
\end{equation}
For the elementary fermions presently observed,
\begin{equation}
\tilde r=r+1.
\end{equation}
The quantum state D$^j_{mm^\prime}$ may be described either as a knotted field ($N,w,\tilde r$) composed of preonic flux tubes according to (5.12)-(5.14), or as a composite particle $(t,t_3,t_0)$ composed of fermionic particles according to (5.15)-(5.17).

\begin{figure}[p]
\centering 
\vspace{-0.25cm}
\mbox{\textbf{\ul{Figure 5.1}}}\\[2ex]
\begin{center}
\begin{tabular}{cc|cc}
 & \underline{$\left( w, r, o \right)$} & & \underline{$\left( w, r, o \right)$} \\
 Leptons, $\mbox{D}^{\frac{3}{2}}_{\frac{3}{2} \frac{3}{2}} \sim a^3$ & & $a$-preons, $\mbox{D}^{\frac{1}{2}}_{\frac{1}{2} \frac{1}{2}}$ \\
 
 \includegraphics[scale=0.42]{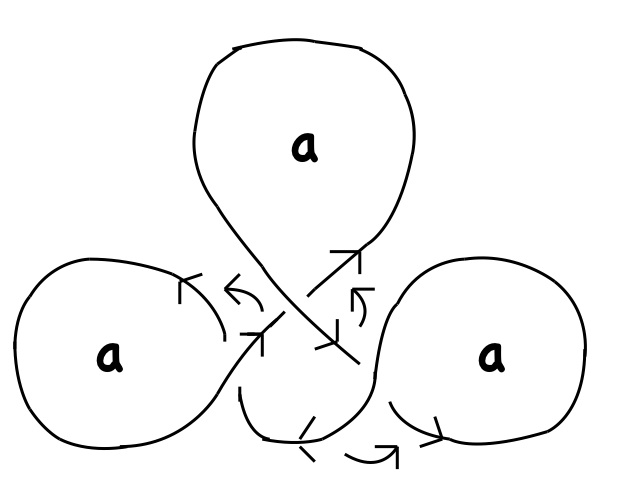} & $\left(3,2,1\right)$ & \includegraphics[scale=0.42]{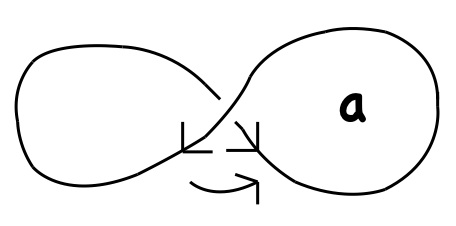} & $\left(1,0,1\right)$ \\
 \hline
  & \underline{$\left( w, r, o \right)$}\\
 Neutrinos, $\mbox{D}^{\frac{3}{2}}_{-\frac{3}{2} \frac{3}{2}} \sim c^3$ & & $c$-preons, $\mbox{D}^{\frac{1}{2}}_{-\frac{1}{2} \frac{1}{2}}$ & \\
 
 \includegraphics[scale=0.420]{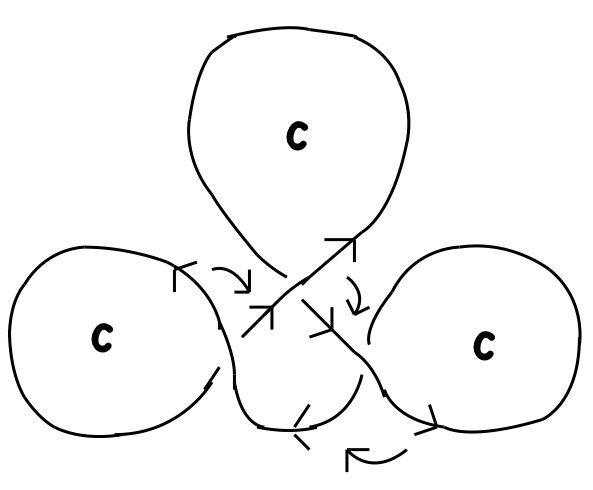} & $\left(-3,2,1 \right)$ & \includegraphics[scale=0.42]{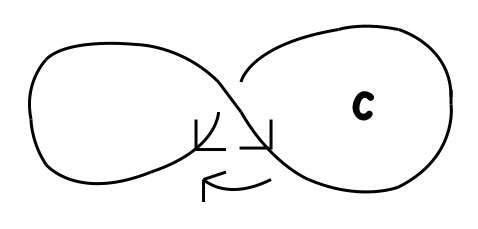} & $\left(-1,0,1 \right)$\\
 \hline
 $d$-quarks, $\mbox{D}^{\frac{3}{2}}_{\frac{3}{2} -\frac{1}{2}} \sim ab^2$ & & $b$-preons, $\mbox{D}^{\frac{1}{2}}_{\frac{1}{2}-\frac{1}{2}}$ & \\
 
\includegraphics[scale=0.420]{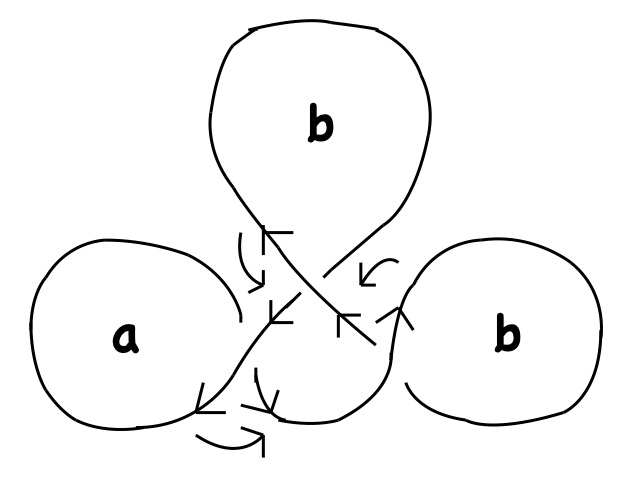} & $\left( 3,-2,1\right)$ & \includegraphics[scale=0.42]{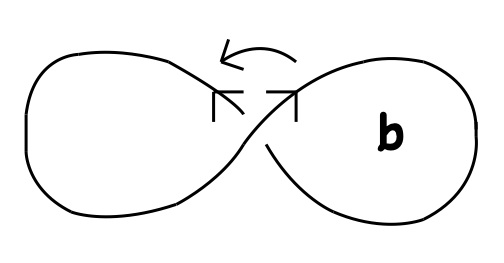} & $\left(1,0,-1\right)$ \\
\hline
$u$-quarks, $\mbox{D}^{\frac{3}{2}}_{-\frac{3}{2} -\frac{1}{2}} \sim cd^2$ & & $d$-preons, $\mbox{D}^{\frac{1}{2}}_{-\frac{1}{2} -\frac{1}{2}}$ \\

\includegraphics[scale=0.42]{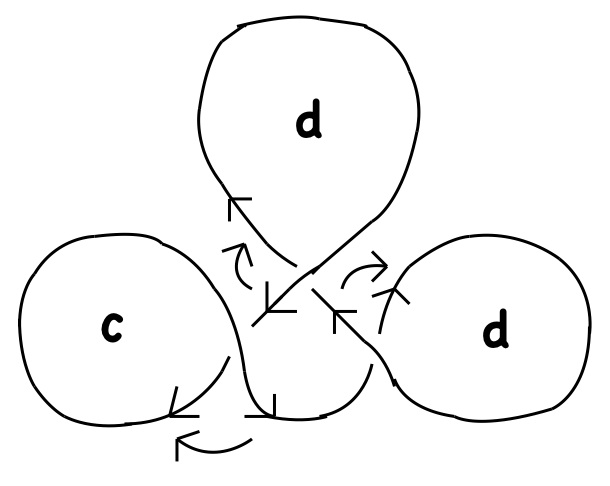} & $\left(-3,-2,1 \right)$ & \includegraphics[scale=0.42]{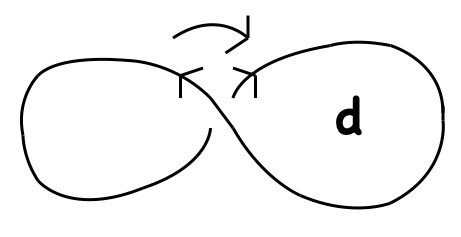} & $\left(-1,0,-1 \right)$ \\
\end{tabular}
\begin{align*}
Q&=-\frac e 6 (w+r+o)\\
(j,m,m^\prime)&=\frac 1 2 (N,w,r+o)
\end{align*}
\end{center}
\end{figure}

The representation of the four trefoils as composed of three overlapping preon loops is shown in Figure 5.1. In interpreting Figure 5.1, note that the two lobes of all the preons make opposite contributes to the rotation, $r$, so that the total rotation of each preon vanishes. When the three $a$-preons and $c$-preons are combined to form leptons and neutrinos, respectively, each of the three labelled circuits is counterclockwise and contributes +1 to the rotation while the single unlabeled shared (overlapping) circuit is clockwise and contributes $-$1 to the rotation so that the total $r$ for both leptons and neutrinos is +2. For the quarks the three labelled loops contribute $-$1 and the shared loop +1 so that $r=-2$.

Equation (5.6) states that the total number of preons, $N^\prime$, equals the number of crossings, ($N$). Since we assume that the preons are fermions, the knot describes a fermion or a boson depending on whether the number of crossings in odd or even.

Viewed as a knot, a fermion becomes a boson when the number of crossings is changed by attaching or removing a curl. This picture is consistent with the view of a curl as an opened preon loop.

Since $a$ and $d$ are antiparticles with opposite charge and hypercharge, while $b$ and $c$ are neutral antiparticles with opposite values of the hypercharge, we may introduce the preon numbers
\begin{align}
\nu_a&=n_a-n_d\\
\nu_b&=n_b-n_c.
\end{align}
Then (5.7) and (5.8) may be rewritten as
\begin{align}
\nu_a+\nu_b&=w\mbox{ }(=-6t_3)\\
\nu_a-\nu_b&=\tilde r\mbox{ }(=-6t_0)
\end{align}
By (5.22) and (5.23) the conservation of the preon numbers and of the charge and hypercharge is equivalent to the conservation of the writhe and rotation, which are topologically conserved at the classical level. In this respect, these quantum conservation laws correspond to the classical conservation laws.

One may view the symmetry of an elementary particle, defined by representations of the SLq(2) algebra, in any of the following ways:
\begin{equation}
\mbox D^j_{mm^\prime}=\mbox D^{3t}_{-3t_3-3t_0}=\mbox D^{\frac N 2}_{\frac w 2 \frac {\tilde r} 2}=\tilde{\mbox D}^{N^\prime}_{\nu_a\nu_b},
\end{equation}
where $N^\prime$ is the total number of preons. The quantum knot-preon complementary representations are related by
\begin{equation}
\tilde{\mbox D}^{N^\prime}_{\nu_a\nu_b}=\sum_{N,w,r}\delta(N^\prime,N)\delta(\nu_a+\nu_b,w)\delta\left(\nu_a-\nu_b,\tilde r\right)\mbox D^{\frac N 2}_{\frac w 2 \frac {\tilde r} 2}
\end{equation}

Since one may interpret the elements $(a,b,c,d)$ of the SLq(2) algebra as creation operators for either preonic particles or flux loops, the D$^j_{mp}$ may be interpreted as a creation operator for a composite particle composed of either preonic particles or flux loops. These two complementary views of the same particle may be reconciled as describing $N$-body systems bound by a knotted field having $N$-crossings as illustrated in Figure 5.2 for $N=3$. In the limit where the three outside lobes become infinitesimal compared to the central circuit, the resultant structure will resemble a three particle system tied together by a Nambu-like string. Since the topological diagram of Figure 5.2 describes loops that have no size or shape, one needs to introduce an explicit Lagrangian to go further.

\begin{figure}[H]
\centering
\mbox{\textbf{\ul{Figure 5.2}}}\\[2ex]
{\small{
\begin{tabular}{cc|cc}
 & \underline{$\left( w, r, o \right)$} & & \underline{$\left( w, r, o \right)$} \\
 Neutrinos, $\mbox{D}^{\frac{3}{2}}_{-\frac{3}{2} \frac{3}{2}} \sim c^3$ & &Leptons, $\mbox{D}^{\frac{3}{2}}_{\frac{3}{2} \frac{3}{2}} \sim a^3$ \\
 
\includegraphics[scale=0.4]{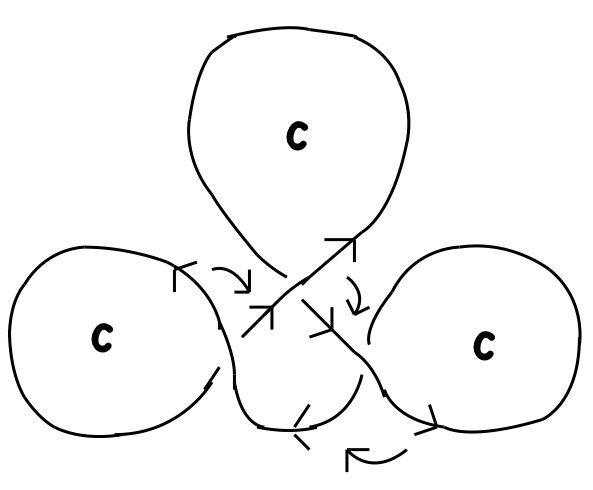} & $\left(1,0,1\right)$ & \includegraphics[scale=0.4]{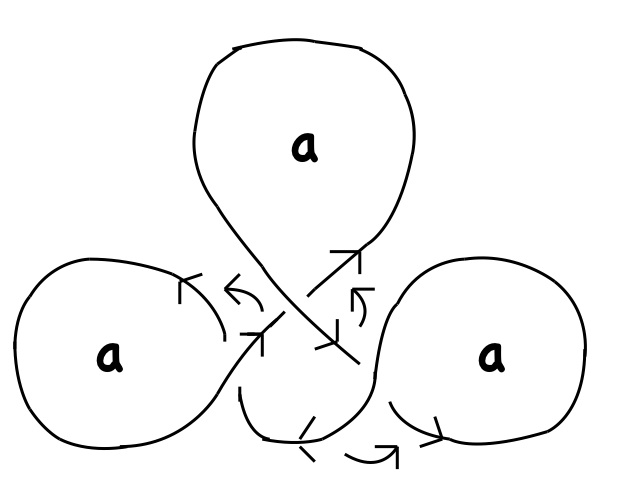} & $\left(3,2,1\right)$  \\
 \hline
 
 \hline
 $d$-quarks, $\mbox{D}^{\frac{3}{2}}_{\frac{3}{2} -\frac{1}{2}} \sim ab^2$ & & $u$-quarks, $\mbox{D}^{\frac{3}{2}}_{-\frac{3}{2} -\frac{1}{2}} \sim cd^2$ \\
 
\includegraphics[scale=0.4]{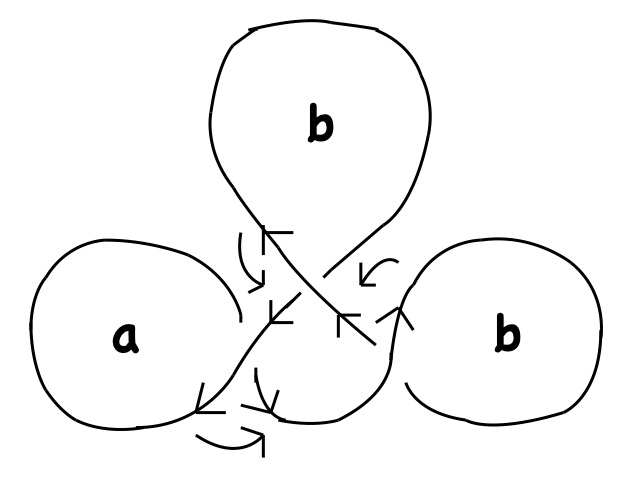} & $\left( 3,-2,1\right)$ & \includegraphics[scale=0.4]{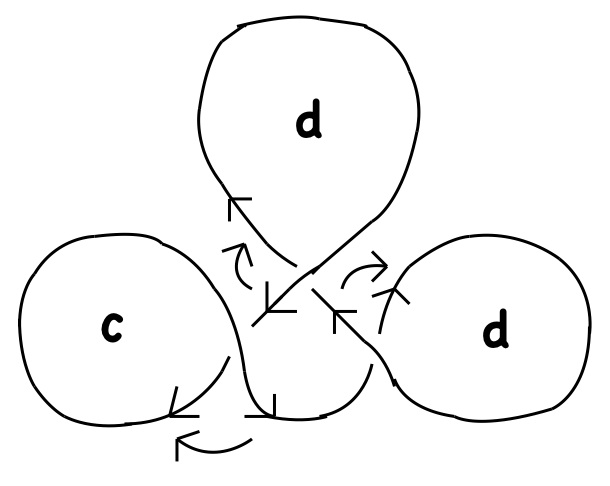} & $\left(-3,-2,1 \right)$\\
\end{tabular}}}

\caption*{The preons conjectured to be present at the crossings are not shown in these figures.}
\end{figure}


\section{The Knot and Preon Lagrangians} 
To construct the knot Lagrangian we replace the left chiral field operators for the elementary fermions, $\Psi_L(x)$, and the electroweak vectors, W$_\mu(x)$, of the standard model by $\Psi_L^{\frac 3 2}(x)$D$^{\frac 3 2}_{mm^\prime}$, and W$_\mu^3(x)$D$^3_{mm^\prime}$, respectively. To construct the preon Lagrangian we replace $\Psi_L(x)$ and W$_\mu(x)$ by $\Psi_L^{\frac 1 2}(x)$D$^{\frac 1 2}_{mm^\prime}$ and W$^1_\mu(x)$D$^1_{mm^\prime}$, respectively. In both the knot and the preon Lagrangians we preserve the local $\mbox{SU(3)}\times\mbox{SU(2)}\times\mbox{U(1)}$ symmetry and therefore the dynamics of the standard model, but the knot factors will introduce form factors in all terms. The factors $\Psi_L^j(x)$ and W$^j_\mu(x)$ in the knot and preon models record the masses and momenta of either the fermions and bosons of the standard model or of the hypothetical preons and preonic vectors. We assume that every right singlet has the same knot factor as the corresponding left triplet. The knot and preon actions obtained according to the above modifications of the standard model have been given\cite{art6}. There the Higgs masses of the standard model get rescaled in the knot model by the factor
\begin{equation}
\langle n | \bar {\mbox D}^{\frac 3 2}_{mm^\prime}\mbox D^{\frac 3 2}_{mm^\prime}|n\rangle
\end{equation}

where the $|n\rangle$ are three eigenstates of the commuting operators $b$ and $c$. The four classes (charged leptons, neutrinos, up and down quarks) have been labelled here by ($m,m^\prime$) in D$^{\frac 3 2}_{mm^\prime}$, and the three members of each class are distinguished by three values of $|n\rangle$ in (6.1).

The electroweak matrix elements are in a similar way rescaled by the form factor
\begin{equation}
\langle n^{\prime\prime} | \bar {\mbox D}^{\frac 3 2}_{m^{\prime\prime}p^{\prime\prime}}\mbox D^3_{mp}\mbox D^{\frac 3 2}_{m^\prime p^\prime}|n^\prime\rangle
\end{equation}
where D$^3_{mp}$ is the knot factor of the electroweak vector and the $|n\rangle$ run over the three mass eigenstates.

The ``flavor states" $|i\rangle$, which we take to be eigenstates of the lowering operator $a$, are coherent sums of the mass states
\begin{equation}
|i\rangle=\sum |n\rangle\langle n|i\rangle
\end{equation}
and a CKM-like matrix
\begin{equation}
\langle i^{\prime\prime} | \bar {\mbox D}^{\frac 3 2}_{m^{\prime\prime}p^{\prime\prime}}\mbox D^3_{mp}\mbox D^{\frac 3 2}_{m^\prime p^\prime}|i^\prime\rangle
\end{equation}
may be parametrized by $q$ and the eigenvalues of $b$ and $c$ on the ground state of the algebra.

In a similar way the Higgs masses of the preons get rescaled by the factor
\begin{equation}
\langle 0| \bar {\mbox D}^{\frac1 2}_{mm^\prime}\mbox D^{\frac 1 2}_{mm^\prime}|0\rangle
\end{equation}
where $j=\frac 1 2$, and $|n\rangle$ is replaced by $|0\rangle$, the ground state of the algebra.

Instead of the matrix (6.4) we consider only the following form factor for the preon-preon interaction as mediated by the preonic vector:
\begin{equation}
\langle 0| \bar {\mbox D}^{\frac1 2}_{m^{\prime\prime}p^{\prime\prime}}\mbox D^1_{mp}\mbox D^{\frac 1 2}_{m^\prime p^\prime}|0\rangle
\end{equation}
The rescaling factors (6.5) and (6.6) may also be parametrized by $q$ and $\beta$ and $\gamma$, the eigenvalues of $b$ and $c$ on $|0\rangle$.


\section{Composite Leptons, Neutrinos, and Quarks}

Although the masses and interactions of the composite leptons, neutrinos, and quarks can be expressed in terms of the knot parameters, $q$, $\beta$, $\gamma$, it may be possible to obtain a more detailed description of the these 12 particles as the three preon structures that are schematically pictured as in Figure 7.1.

\begin{figure}[ht]\centering
\centering \mbox{\textbf{\ul{Figure 7.1}}}\\[2ex]
~~~Leptons \qquad\qquad ~~Neutrinos \qquad\qquad Down Quarks \qquad\qquad Up Quarks\\
\scalebox{0.41}{\includegraphics{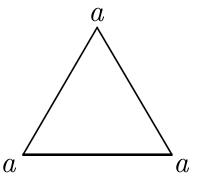}}\qquad \scalebox{0.41}{\includegraphics{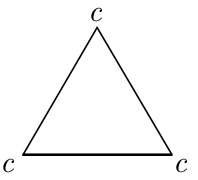}} \qquad\scalebox{0.41}{\includegraphics{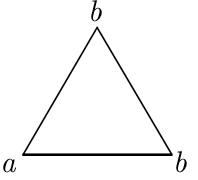}}\qquad \scalebox{0.41}{\includegraphics{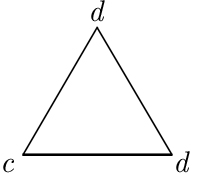}}\\
$a^3$ \qquad\qquad\qquad\qquad $c^3$ \qquad\qquad\qquad\qquad ~$ab^2$ \qquad\qquad\qquad\qquad $cd^2$
\end{figure}

The topological diagrams in Figure 5.2 may be shrunk into the effectively triangular shapes in Figure 7.1 by reducing the outside loops of the trefoils to infinitesimal loops. Then the two body forces result from the matrix elements that connect the fermionic preons and are mediated by the preonic vectors. The relevant two body bonds are
\begin{align*}
\begin{aligned}[c]
\mbox{leptons: } &a-a \\
\mbox{down quarks: } &a-b \mbox{ and } b-b 
\end{aligned}
\qquad\qquad
\begin{aligned}[c]
\mbox{neutrinos: } &c-c \\
\mbox{up quarks: } &c-d \mbox{ and } d-d
\end{aligned}
\end{align*}
Then we have Table 7.1 describing form factors for the two body forces.
\begin{table}[H]
\centering
\small \setlength{\tabcolsep}{13pt}
\renewcommand{\arraystretch}{2}
\begin{tabular}{l | c c l}
\multicolumn{4}{c}{\ul{\textbf{Table 7.1}}}\\[1ex]\hline\hline
& D$^{\frac 3 2}_{mp}$ & Bond & \multicolumn{1}{c}{Form Factor} \\ \hline
leptons & $a^3$ & $a-a$ & $\bar{\mbox D}^{\frac 1 2}_{\frac 1 2 \frac 1 2}$ D$^1_{00}$ D$^{\frac 1 2}_{\frac 1 2 \frac 1 2} = \bar a (ad+bc)a$ \\
neutrinos & $c^3$ & $c-c$ & $\bar{\mbox D}^{\frac 1 2}_{-\frac 1 2 \frac 1 2}$ D$^1_{00}$ D$^{\frac 1 2}_{-\frac 1 2 \frac 1 2} = \bar c (ad+bc)c$\\
down quarks & $ab^2$ & $a-b$ & $\bar{\mbox D}^{\frac 1 2}_{\frac 1 2 \frac 1 2}$ D$^1_{01}$ D$^{\frac 1 2}_{\frac 1 2 -\frac 1 2} = \bar a (ac)b$\\
& & $b-b$ & $\bar{\mbox D}^{\frac 1 2}_{\frac 1 2 -\frac 1 2}$ D$^1_{00}$ D$^{\frac 1 2}_{\frac 1 2 -\frac 1 2} = \bar b (ad+bc)b$\\
up quarks & $cd^2$ & $c-d$ & $\bar{\mbox D}^{\frac 1 2}_{-\frac 1 2 \frac 1 2}$ D$^1_{01}$ D$^{\frac 1 2}_{-\frac 1 2 -\frac 1 2} = \bar c(ac)d$\\
& & $d-d$ & $\bar{\mbox D}^{\frac 1 2}_{-\frac 1 2 -\frac 1 2}$ D$^1_{00}$ D$^{\frac 1 2}_{-\frac 1 2 -\frac 1 2} = \bar d (ad+bc)d$\\[1ex]\hline\hline
\end{tabular}
\end{table}

\noindent The operator form factors may all be reduced by the algebra (A) to functions of $bc$ and $q$. If one then reduces these operators by setting
\begin{align}
d&=\bar a\\
c &= -q_1\bar b
\end{align}
and thereby replacing SLq(2) by SUq(2), one finds that the strength of these bonds depends on the values of $q$ and $\bar bb$.

Within the SLq(2) kinematics there are several options in constructing an effective Hamiltonian for the three body structures that represent the leptons, neutrinos, and quarks. These possible three body Hamiltonians permit electric, magnetic, gluon, and gravitational forces, but the model, as here described, only allows these forces as mediated by the preonic adjoint field and therefore proportional to the form factors in Table 7.1. A successful model must have an angular momentum of $\frac h 2$ and should also have only three bound states, corresponding to the three generations. The quanta of the binding field lie in the adjoint representation with wave functions that are represented by normal modes of $w^+_\mu(x)$D$^1_{-1 0}$, $w^-_\mu(x)$D$^1_{1 0}$, $z_\mu(x)$D$^1_{0 0}$, and $a_\mu(x)$D$^1_{0 0}$.

The preonic photon field $a_\mu$D$^1_{00}$ produces a Coulombic potential while the charged $w$ particles and the neutral $z$, being massive, are responsible for a Yukawa-type potential. The range of the Yukawa potentials is determined by the masses of the charged $w$ and the neutral $z$. The Higgs masses of these particles are in turn determined by the vacuum expectation values of the Higgs scalars and are rescaled by
\begin{equation}
\langle 0|\bar{\mbox D}^1_{mm^\prime}\mbox{ D}^1_{mm^\prime}|0\rangle.
\end{equation}
All of these vacuum expectation values are dependent on $q$, $\beta$, and $\gamma$. Similarly the form factors, rescaling the interactions, are dependent on the same parameters. One can construct a formal effective Hamiltonian dependent on the parameters $q$, $\beta$, and $\gamma$.

In order to achieve an adequately strong binding at very short range, 
it is important that the parameters ($q,\beta,\gamma$), determining the strength of the form factors, and the factor coming from the Higgs scalar, determining the range of the Yukawa potentials, be themselves sufficiently large. The magnitude of these parameters in turn depends on their physical meaning, which we now briefly consider.


\section{A Possible Physical Interpretation of the SLq(2) \\Algebra and of the Deformation Parameter $\mathbf{q}$}
In an earlier work\cite{art5}, an implicit connection between the SLq(2) algebra and the 2d projections of the classical 3d-knots was made through the matrix
\begin{equation}
\varepsilon_q=\left(\def\arraystretch{1} \begin{array}{c c}
0 & q^{-\frac 1 2} \\
-q^{\frac 1 2} & 0 \end{array} \right)
\end{equation}
which is invariant under the following transformation
\begin{equation}
\mbox T\varepsilon_q \mbox T^t=\mbox T^t\varepsilon_q \mbox T=\varepsilon_q,
\end{equation}
where the elements of T define the SLq(2) algebra and where $\varepsilon_q$ underlies the Kauffman algorithm for associating the Kauffman polynomial with a knot \cite{art5}.

\noindent In Equation (2.2) of the present paper, $\varepsilon_q$ is replaced by 
\begin{equation}
\varepsilon=\left( \def\arraystretch{1}\begin{array}{c c}
0 & \alpha_2 \\
-\alpha_1 & 0 \end{array} \right).
\end{equation}
Then the SLq(2) algebra (A) is again generated by (8.2) but with
\begin{equation}
q=\frac{\alpha_1}{\alpha_2}.
\end{equation}
If one further imposes
\begin{equation}
\det \varepsilon=1
\end{equation}
then
\begin{equation}
\alpha_1\alpha_2=1
\end{equation}
and
\begin{equation}
\varepsilon=\varepsilon_q.
\end{equation}

Then (8.3), restricted by (8.5), is equivalent to (8.1) and the knot model may be based on either $\varepsilon$ or $\varepsilon_q$. By taking advantage of the fact that $\varepsilon$ is a two-parameter matrix while $\varepsilon_q$ depends on only a single parameter, however, one may describe a wider class of physical theories with $\varepsilon$. If the physical situation that the theory is being asked to describe is characterized by two interacting gauge fields, with two charges, $g$ and $g^\prime$, on the same particle, one may attempt to give physical meaning to $q$, and gain a possible improvement in the model by embedding $g$ and $g^\prime$ in $\varepsilon$ as follows
\begin{equation}
\varepsilon=\left(\def\arraystretch{1} \begin{array}{c c}
0 & \frac{g(E)}{\sqrt{\hbar c}} \\
-\frac{g^\prime(E)}{\sqrt{\hbar c}} & 0 \end{array} \right),
\end{equation}
where $g(E)$ and $g^\prime(E)$ are energy dependent coupling constants that have been normalized to agree with experiment at hadronic energies.
\\Then $q$ is defined by (8.8) and (8.4) as
\begin{equation}
q(E)=\frac{g^\prime(E)}{g(E)}.
\end{equation}
If (8.5) is also imposed, then
\begin{equation}
g(E)g^\prime(E)=\hbar c,
\end{equation}
which is like the Dirac restriction on magnetic poles:
\begin{equation}
eg=\hbar c.
\end{equation}

In the electroweak knot model it is argued that the electroweak experimental data suggest an SLq(2) extension of the standard model. To the extent that this view is correct it appears that the sources of the electroweak field are knotted, but the possible physical origins of the additional ``knot" degrees of freedom have not been identified. A possible origin of the ``knotting" is the deformation of the electroweak SU(2)$\times$U(1) structure by SU(3). Since the leptons and neutrinos, appearing as $a^3$ and $c^3$ particles in the SLq(2) model, have already been given SU(3) indices to protect the Pauli principle, the gluon field is implicit in this model and a possible interpretation of (8.8) is then $(g^\prime,g)=(e,g)$ or $(g,e)$, where $g$ is the gluon charge, and $e$ is an electroweak coupling constant.

Then (8.10) would become
\begin{equation}
eg=\hbar c,
\end{equation}

Since $g$ and $e$ are running coupling constants, the SLq(2) parameter $q$, which is either $\frac e g$ or $\frac g e$, is also a running and dimensionless coupling constant. If $e$ increases with energy and $g$ decreases with energy according to asymptotic freedom, $q$ may become very large or very small at the high energies where the interaction and mass terms become relevant for fixing the three particle bound states representing charged leptons, neutrinos, and quarks. 
Although there is currently no experimental data suggesting the interpretation of $q$ as the ratio of an $e$ and a $g$,
such a relation (resulting from a possible physical interpretation of the otherwise undefined matrix $\varepsilon$ in (8.8)) could be explored 
since $e$, $g$ and $q$ can be independently measured.


\section{Bound Preons}
A major uncertainty in these realizations of the knot model lies in the unknown values of the Higgs factors and more fundamentally in the nature of the Higgs fields and their relation to the gravitational field.

It is not possible to construct a more predictive SLq(2) modification of the standard 
model until the Higgs factor, as well as $q$ and $\beta$, are understood. Finally, the models discussed here resemble familiar composite particles like H$^3$, but it is possible that the H$^3$ example is not appropriate and that the preons are always bound. In this case the preons may not have an independent existence but may be particular field structures, or elements of larger field structures, carrying no independent degrees of freedom. In the SLq(2) model described here the elementary fermions are three-preon composite particles bound by a trefoil field structure. It is possible to assume that the trefoil field structure is a trefoil flux tube carrying energy, momentum, and charge and that energy, momentum, and charge are concentrated at the three crossings. It is then possible to regard these three concentrations of energy, momentum, and charge at the three crossings as actually defining the three preons without postulating their independent existence with independent degrees of freedom. Since the number of preons in any composite particle is always equal, in the SLq(2) model, to the number of crossings (by (5.6)), this view of the preons as tiny solitonic regions of field surrounding the crossings holds for all composite particles considered here. This view of the elementary particles as lumps of field is sometimes described as a unitary field theory and has also been examined in other solitonic contexts.

\section*{Acknowledgements}

I thank A.C. Cadavid and J. Smit for helpful discussions.

\newpage

\bibliographystyle{unsrt}
\bibliography{finkelsteinbib}

\begin{thebibliography}{1}

\bibitem{art1}
H.~Harari.
\newblock {\em Physics Letters B}, 86:83--86, 1979.

\bibitem{art2}
M.~Shupe.
\newblock {\em Physics Letters B}, 86:87--92, 1979.

\bibitem{art3}
R.~Finkelstein.
\newblock ar{X}iv:1205.1026v3 [hep-th].

\bibitem{art4}
H.~Harari and N.~Seiberg.
\newblock {\em Nuclear Physics B}, 204:141--167, 1982.

\bibitem{art5}
R.~Finkelstein.
\newblock ar{X}iv:1011.2545v1 [hep-th].

\bibitem{art6}
R.~Finkelstein.
\newblock ar{X}iv:1301.6440v3 [hep-th].

\end{thebibliography}

\end{document}